\documentclass[5p]{elsarticle}
\usepackage{amsfonts}
\usepackage{amssymb}
\usepackage{array}
\usepackage[cmex10]{amsmath}
\usepackage{lineno,hyperref}
\modulolinenumbers[5]

\journal{NeuroImage}

\begin{document}
	
	\begin{frontmatter}
		
		\title{Scalable Algorithms for Generating and Analyzing Structural Brain Networks with a Varying Number of Nodes}

		\author{Yu Jin\textsuperscript1, Joseph F. JaJa\textsuperscript1, Rong Chen\textsuperscript2, Edward H. Herskovits\textsuperscript2}
		\address{\textsuperscript1Institute for Advanced Computer Studies} \address{Department of Electrical and Computer Engineering}
		\address{University of Maryland, College Park, USA}
	
		\address{\textsuperscript2Department of Radiology}
		\address{University of Maryland, Baltimore, USA}
			
		\begin{abstract}
			Diffusion Magnetic Resonance Imaging (MRI) exploits the anisotropic diffusion of water molecules in the brain to enable the estimation of the brain's anatomical fiber tracts at a relatively high resolution. In particular, tractographic methods can be used to generate  whole-brain anatomical connectivity matrix where each element provides an estimate of the connectivity strength between the corresponding voxels. Structural brain networks are built using the connectivity information and a predefined brain parcellation, where the nodes of the network represent the brain regions and the edge weights capture the connectivity strengths between the corresponding brain regions. This paper introduces a number of novel scalable methods to generate and analyze structural brain networks with a varying number of nodes. In particular, we introduce a new parallel algorithm to quickly generate large scale  connectivity-based parcellations for which voxels in a region possess highly similar connectivity patterns to the rest of the regions. 
		We show that the corresponding regional structural consistency 
is always superior to  randomly generated parcellations over a wide range of parcellation sizes.  Corresponding brain networks with a varying 
number of nodes are analyzed using standard graph-theorectic measures, as well as, new measures derived from spectral graph theory. Our results  indicate increasingly more  statistical power of brain networks with larger numbers of nodes and the relatively unique shape of the spectral profile of large brain networks relative to other well-known networks.
			
		\end{abstract}
		
		\begin{keyword}
		Connectivity-based brain parcellation \sep Structural brain network \sep Spectral clustering \sep Network analysis \sep Spectral analysis
			
		\end{keyword}
		
	\end{frontmatter}
	
	
	\section{Introduction}
	Non-invasive neuroimaging technologies, combined with computational techniques, have played a central role in providing critical insights into the structural and functional organization of the brain. Corresponding brain networks have been extensively studied using different notions of connectivity such as structural connectivity, based on synapses between neighboring neurons or fiber tracts between brain regions; functional connectivity based on statistical correlations of activation levels between distinct regions; or effective connectivity that captures causal interactions that can be inferred from network perturbations or time-series analyses. These networks have been studied using segregation measures that refer to the degree to which a network's nodes form separate cliques or clusters, and integration measures that refer to the capacity of the network as a whole to become interconnected and exchange information \cite{bullmore2009complex}.

	\par This work focuses on structural brain networks derived from Diffusion Tensor Imaging (DTI) data.  DTI allows one to observe the diffusion process of water molecules in brain tissues under magnetic fields with different strengths and along different directions \cite{sporns2013structure, qi2015, herskovits2015edge}. Modern tractographic methods can then be used to estimate the anatomical connectivity information. Using this connectivity information and a set of predefined regions of interests (ROIs) or whole brain parcellations, structural brain networks can be built and analyzed. A significant number of efforts have been devoted to study the topological properties of structural brain networks using graph-theoretic measures such as characteristic path length, clustering coefficient, local and global efficiency, degree distribution, betweenness centrality etc \cite{kaiser2011tutorial, sporns2013structure, zalesky2010whole}. These studies have resulted in a number of interesting findings regarding brain networks such as scale-freeness, small-world property, and modular organization.  Moreover, researchers have explored  the topological differences of the networks corresponding to distinct  population groups \cite{ingalhalikar2014sex, gong2009age, iturria2013anatomical}. It was observed that subjects with certain brain disorders seem to have structural abnormalities in terms of local and global graph properties. Given the early promising results, a number of tutorials and tools have been developed to facilitate  brain network analysis \cite{kaiser2011tutorial, rubinov2010complex}. 
	
	\par As described in the literature, structural brain networks have been constructed based on predefined parcellations that typically use traditional anatomical brain atlases, such as the Automated Anatomical Labeling (AAL) atlas \cite{tzourio2002automated, cao2013probabilistic} and the J\"{u}lich histological atlas \cite{eickhoff2005new}, or some random parcellation\cite{zalesky2010whole}.  Researchers have also developed parcellation algorithms that are  based on well-known clustering methods such as Gaussian Mixture Model (GMM), k-means clustering, hierarchical clustering \cite{moreno2014hierarchical}; these algorithms partition the brain mask into many spatially contiguous regions. These studies have 
	primarily focused on specific anatomical regions of the brain such as the human inferior parietal cortex complex (IPCC) \cite{ruschel2014connectivity}, the lateral parietal cortex \cite{mars2011diffusion}, the temporoparietal junction area (TPJ) \cite{mars2012connectivity}, the dorsal frontal cortex \cite{sallet2013organization}, the ventral frontal cortex \cite{neubert2014comparison}, cingulate and orbitofrontal cortex \cite{neubert2015connectivity}, and Broca's areas \cite{anwander2007connectivity}. Most of these methods 
	are not easy to generalize to determine whole-brain brain parcellations while enforcing the constraint of connectivity-based homogeneous regions. Note 
	that the choice of the parcellation plays a critical role in determining the quality of the corresponding  brain networks. Traditional atlases and random parcellations do not make use of  the connectivity information that can be derived from tractographic methods, which may alter the statistical effectiveness of subsequent structural network analysis. Another important parameter that received relatively little attention in the literature is the  number of nodes in the brain network. In particular, it would be interesting to figure out how the number of nodes affects various network properties and whether larger brain networks capture additional structural patterns than smaller ones.

	\par The graph theoretical analysis is usually performed on sparse brain networks derived from the original relatively dense brain networks \cite{sporns2013structure}. The original  brain networks may contain weak connections that are eliminated by a thresholding process typically based a single threshold value. But this threshold value  affects the structure and sparsity of the brain networks and hence may heavily impact the corresponding graph properties. In fact, often the threshold value is selected to enforce certain sparsity of the graph  without fully 
	understanding the impact on the corresponding topological properties of the derived netowrks. During  the thresholding (and binarizing) process a significant amount of connectivity information is lost. This paper explores the use  spectral graph theory to be introduced later for analyzing brain networks. Spectral analysis is performed on the original weighted brain network and can capture almost all the topological invariants of the network. No threstholding process is needed, and the corresponding spectral analysis seems to be much more robust as we will see later.


	
	\par The main contributions of this paper are the following.
	\begin{itemize}
		\item Development of a fast parallel algorithm to generate whole-brain connectivity-based parcellations with a very large number of regions. 
		The resulting regions are shown to have a very good degree of structural consistency in terms of the corresponding brain networks.  
		\item Analysis of structural brain networks for distinct population groups using a varying number of nodes, up to $500$ nodes, which indicates richer 
		structural patterns with larger numbers of nodes.
		\item Preliminary indication of the robustness and the power of network parameters derived form spectral graph theory when 
		compared to the traditional graph-theoretic measures. In particular, the shape of 
		the spectrum of such networks seems to be unique among the spectrums of other well-known networks.
	\end{itemize}
	
	\par The rest of the paper is organized as follows. Section 2 introduces the dataset used in this study as well as the software tools used to generate the connectivity information. Section~3 briefly describes the GPU-CPU heterogeneous system and necessary software libraries used for our parallel  implementation. Section~4 provides a brief overview of the parallel parcellation algorithm while Section~5 covers the evaluation of the connectivity-based parcellations and the network analysis using both the traditional graph-theoretic measures and new parameters 
	derived from spectral graph theory. We conclude the work in Section~6.

	\section{Materials}
	\subsection{Diffusion MRI Dataset}
	The diffusion MRI dataset used in this research is taken from the publicly available Nathan Kline Institute (NKI)/ Rockland dataset\footnote{\href{http://fcon_1000.projects.nitrc.org/indi/pro/nki.html}{http://fcon\_1000.projects.nitrc.org/indi/pro/nki.html}}. The neuroimage sample consists of data for individuals whose ages range from 4 to 85. The demographics of our sample are shown in Table \ref{Demographics}. A comprehensive psychiatric interviews and behavioral assessments are included to help explore the brain-behavior relationship. The diffusion MRI was performed using a SIEMENS MAGNETOM TrioTim syngo MR B15 system. The high-angular resolution diffusion imaging protocol was used to assess white matter integrity as measured by fractional anisotropy. Diffusion tensor data were collected using a single- shot, echo-planar, single refocusing spin-echo, T2-weighted sequence with a spatial resolution of 2.0$\times$2.0$\times$2.0mm. The sequence parameters were:\\ TE/TR=91/10000ms, FOV=256mm, axial slice orientation with 58 slices, 64 isotropically distributed diffusion weighted directions, two diffusion weighting values (b=0 and 1000s/mm$^{2}$) and six b=0 images. These parameters were calculated using an optimization technique that maximizes the contrast to noise ratio for FA measurements. For each subject, the image data consists of 76 volumes of 3D images of dimensions 128$\times$128$\times$53, each voxel representing 2.0mm$\times$2.0mm$\times$2.0mm brain volume.
	\begin{table*}[!hbt]
		\caption{Demographics of NKI Sample}
		\label{Demographics}
		\centering
		\begin{tabular}{|c|c|c|c|c|c|c|c|c|c|}
			\hline
			\textbf{Age Group} & \textbf{4-9} & \textbf{10-19} & \textbf{20-29} & \textbf{30-39} & \textbf{40-49} & \textbf{50-59} & \textbf{60-69} & \textbf{70-85} & \textbf{Total} \\
			\hline
			\textbf{Female} & 4 & 18 & 19 & 8 & 5 & 6 & 9 & 10 & 79 \\
			\hline
			\textbf{Male} & 5 & 16 & 24 & 9 & 26 & 8 & 5 & 4 & 97\\
			\hline
			\textbf{Total} & 9 & 34 & 43 & 17 & 31 & 14 & 14 & 14 & 176 \\
			\hline
		\end{tabular}
	\end{table*}

	\section{Computational Environment}
	\subsection{CPU-GPU Heterogeneous Platform}
	We use a heterogeneous CPU-GPU platform to generate and analyze our structural brain networks. Standard heterogeneous platforms consist of a multi-core CPU and one or more general-purpose GPU (GPGPU). The CPU and GPU communicate through the Peripheral Component Interconnect Express (PCIe) bus. The multi-core CPU is programmed as a shared memory system with a large main memory and multiple levels of cache. Most existing software packages are implemented based on single or multi-core CPU architecture. On the other hand, GPUs typically have thousands of cores that can execute data parallel operations extremely fast. They have been used extensively for all types of computations especially 
	scientific computations. However programming a GPU is typically a non-trivial task. 
	Since the CPUs and GPUs have somewhat complementary strengths, modern applications try to combine them in 
	single platforms \cite{mittal2015survey}. Hybrid computing environments, which collaboratively combine the computational advantages of GPUs and CPUs, further boost the overall performance \cite{lee2014boosting, agulleiro2012hybrid, wu2014optimized, wuachieving, lihybrid, akimova2012parallel}.  However such a performance gain is typically achieved at the expense of a significant additional programming complexity. The details of the platform used in this study are shown in Table \ref{Specifics}.
	\begin{table}[hbt]
		\caption{System Details}
		\label{Specifics}
		\centering
		\begin{tabular}{|c|c|}
			\hline
			CPU Model & Intel Xeon E5-2690 \\
			\hline 
			CPU Cores & 8 \\
			\hline 
			DRAM Size & 128GB \\
			\hline 
			GPU Model & Tesla K20c\\
			\hline 
			Device Memory Size & 5GB GDDR5 \\
			\hline
			SMs and SPs & 13 and 192\\
			\hline 
			Compute Capability & 3.5\\
			\hline
			CUDA SDK & 7.5 \\
			\hline 
			PCIe Bus & PCIe x16 Gen2 \\
			\hline
		\end{tabular}
	\end{table}
	
	\subsection{Library Dependencies}
	Our brain parcellation scheme is based on a fast parallel implementation of the spectral clustering algorithm described in \cite{yu2016spectral}. We leverage the following libraries to implement our brain parcellation scheme on CPU-GPU heterogeneous platforms.
	\subsubsection{CUDA Library}
\begin{figure}[!t]
		\centering
		\includegraphics[width = \hsize]{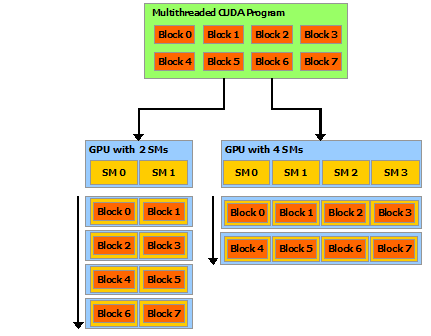}
		\caption{CUDA Program Model}
		\label{cuda}
	\end{figure}

	The Compute Unified Device Architecture (CUDA) is a general-purpose multithreaded programming model that utilizes the large number of GPU cores to solve complex computational problems. The CUDA programming model assumes a heterogeneous system with a host CPU and one or more GPUs as co-processors. Each GPU has an array of Streaming Multiprocessors (\emph{SM}), each of which has a number of Streaming Processors (\emph{SP}) that execute instructions concurrently. The parallel computation on GPU is invoked by calling customized kernel functions using thousands of threads. The kernel function is executed by blocks of threads independently. Each block of threads can be scheduled on any Streaming Multiprocessor (SP) as shown in Figure \ref{cuda}. The kernel function takes as  parameters the number of blocks and the number of threads within a block.
	
	\par In addition, we use the efficient BLAS libraries for both sparse\footnote{\href{http://docs.nvidia.com/cuda/cusparse/}{http://docs.nvidia.com/cuda/cusparse/}} and dense\footnote{\href{http://docs.nvidia.com/cuda/cublas/}{http://docs.nvidia.com/cuda/cublas/}} matrix computations. Our implementation also relies on the Thrust library, which resembles the C++ Standard Template Library (STL) that significantly improves productivity.
	\begin{figure*}[hbt!]
		\centering
		\includegraphics[width = 0.8\hsize]{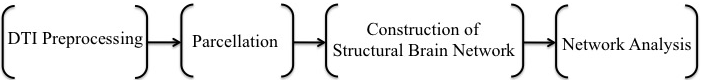}
		\caption{Overview of the Generation and Analysis of  Structural Brain Networks}
		\label{overview}
	\end{figure*}
	\subsection{ARPACK Software}
	ARPACK is a software package designed to solve large-scale eigenvalue problems \cite{lehoucq1997arpack}. It contains highly optimized Fortran subroutines that solve symmetric, non-symmetric and generalized eigenvalue problems which are widely used in modern scientific software packages  such as Matlab and Python scientific packages. ARPACK is based on the Implicitly Restarted Arnoldi Method (IRAM) with non-trivial numerical optimization techniques \cite{lehoucq1996deflation, sorensen1992implicit}. The eigenvalue problem is efficiently solved by collaboratively combining the interfaces of ARPACK and cuSPARSE library.
	
	\section{Methods}
	An overview of the workflow for generating and analyzing structural brain networks is illustrated in Figure~\ref{overview}.  . The DTI preprocessing step extracts the connectivity information represented as a large-scale connectivity matrix by using 
probabilistic tractography. A scalable parallel algorithm is then used to generate parcellatioins  with up to 500 regions  \cite{yu2016spectral,jin2015}. Structural brain networks are constructed from the parcellations and 
the connectivity information. Finally we apply both  traditional graph-theoretical methods and new methods based on spectral 
grpah theory to analyze the brain networks. The details of each step are described below.
	
	\subsection{Dataset Preprocessing}
	\subsubsection{Nonlinear Registration}
	The diffusion images of all subjects are registered to Montreal Neurological Institute (MNI) standard space using the nonlinear registration package \emph{FNIRT} in FSL software \cite{andersson2007non}. The nonlinear registration process generates the warping coefficients that balance the similarity between the diffusion image and the standard MNI152 image, and the smoothness of the warping coefficients.
	\begin{figure*}[hbt!]
		\centering
		\includegraphics[width = 0.8\hsize]{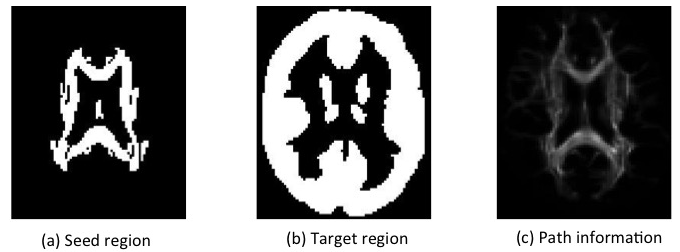}
		\caption{Left: Seed region which is JHU white matter atlas. Middle: target region which is the AAL mask. Right: the result of probabilistic tractography which models the distribution of neuron fiber bundles where the intensity of each voxel represents the number of streamlines passing through that voxel. All figures are imaged in the axial plane}
		\label{tractography}
	\end{figure*}
	
	\subsubsection{Probabilistic Tractography}
	The probabilistic tractography method is one of the modern tractography methods which has been used extensively to extract anatomical connectivity information. For each voxel, the diffusion tensor is built to model the fiber distribution through the \emph{BEDPOSTX} package in FSL \cite{behrens2007probabilistic}. The probabilistic tractography is then processed through the probabilistic tractography toolbox in FSL \cite{behrens2003characterization} by sending out streamlines from seed regions and propagate through target regions. The JHU DTI-based white matter atlas\footnote{\href{http://cmrm.med.jhmi.edu/}{http://cmrm.med.jhmi.edu/}} is specified as the seed region. The AAL mask is specified as the target region, which is also considered as our region of interests to be parcellated. We generate 500 streamlines from every voxel in a seed region. These streamlines are propagated following the cross fiber distribution derived from the voxel-level diffusion tensors. Curvature threshold is enforced to eliminate unqualified streamlines. The distance correction option is set to correct for the fact that the distribution drops as travel distance increases. The tractography output is a structural connectivity network modeled as a large weighted graph where each node is a voxel in the target region space and each edge weight corresponds to relative connectivity strength in terms of the number of streamlines connecting the corresponding pair of voxels. Figure \ref{tractography} shows the seed and target regions as well as the tractography result of a subject.  
	\subsection{Brain Parcellation Scheme}
	The brain parcellation scheme takes as input the probabilistic tractography results  represented as a large connectivity matrix. The number of voxels in the AAL mask is $155,794$ and the connectivity matrix is a sparse matrix of size $155,794 \times 155,794$. Given a positive integer value $k$, the brain parcellation problem is to segment the brain's grey matter into $k$ spatially contiguous regions, such that the connectivity profiles of the voxels in each region are as similar as possible. Moreover, we expect the parcellations to be stable and reproducible, as well as, consistent among members of a structurally 
homogeneous population sample.. 
	
	We next introduce our notion of a \emph{connectivity profile} followed by an overview of the methods used in our parcellation algorithm.  
	
	\begin{figure}[!t]
		\centering
		\includegraphics[width = \hsize]{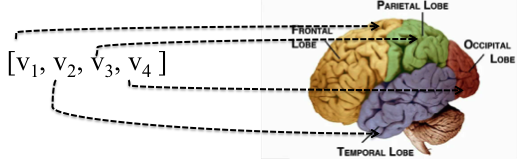}
		\caption{Illustration of connectivity profile for each voxel}
		\label{connectivity}
	\end{figure}
	
	\subsubsection{Connectivity profile}
	For each voxel, the connectivity profile is the signature that discriminates a voxel from the rest of voxels based on connectivity information. Parcellations are built by clustering voxels with similar connectivity profiles together. As shown in Figure \ref{connectivity}, the connectivity profile of a voxel is defined as an array of weights, where each weight represents the  connectivity strength from that voxel to the voxels within each  region defined by a given brain segmentation. As shown in \cite{jin2015}, our parcellation algorithm generates almost identical parcellations regardless of the initial brain segmentation used to define the connectivity profiles as long as the segmentation has a sufficient number 
	of regions. 
	

	\begin{table*}[hbt!]
		\centering
		\begin{tabular}{l}
			\hline \hline
			\textbf{Algorithm 1} Connectivity-based Brain Parcellation Scheme \\
			\hline	
			1. Generate the probabilistic tractography results from diffusion MRI data.\\	
			2.	Initialize a parcellation as a random spatial brain segmentation, to be used to define the connectivity profile of each voxel.\\
			\textbf{Repeat}\\
			3. Define the connectivity profiles using the current brain parcellation.\\
			4. update the spatial similarity graph using the connectivity profiles.\\
			5. Apply spectral clustering algorithm on the current spatial similarity graph to generate the brain parcellation. \\
			6. Measure the similarity between the new parcellation and the previous parcellation used to define connectivity profiles.\\
			\textbf{Until} the similarity measurement exceeds some threshold.\\
			7. Return the parcellation result.\\
			\hline \hline	
		\end{tabular}
	\end{table*}
	
	Our brain parcellation scheme is an iterative algorithm that is  briefly summarized in Algorithm~1, and is explained in detail in \cite{jin2015}. Each major step of the brain parcellation scheme is described next. 
	
	\subsubsection{Spatial similarity graph}
\begin{figure}[!hbt]
		\centering
		\includegraphics[width = .5\hsize]{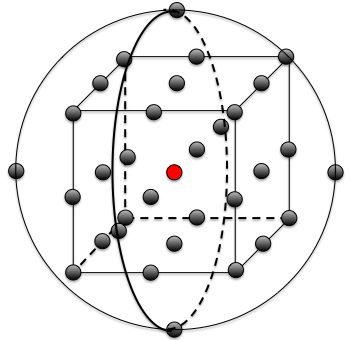}
		\caption{At most 32 neighbors of a voxel are within a sphere of radius $r = 2$.}
		\label{neighbor}
	\end{figure}

	The spatial-constraint similarity graph is formed using spatial adjacency and the connectivity profiles as follows. The voxels define the nodes of the graph. Two nodes are connected by an edge if and only if the corresponding voxels are within a spatial distance of radius $r$. In this study, we use $r = 2$  and hence the number of neighbors of each node is at most $32$ as shown in Figure \ref{neighbor}. Each edge is weighted by the correlation coefficient between the connectivity profiles of its end points.
	
	\subsubsection{Spectral clustering}
	The spectral clustering algorithm is initially proposed to solve minimum graph-cut problems. Specifically, the algorithm partitions a similarity graph into several subgraphs with the objective of minimizing the total weight of the edges connecting the subgraphs subject to a constraint on the relative sizes of the subgraphs \cite{von2007tutorial, ng2002spectral, yu2003multiclass}. In our case, the subgraphs induce a spatially contiguous segmentation of the AAL mask. The algorithm results in a solution where the voxels within the same region have similar connectivity profiles and voxels across different regions have dissimilar connectivity profiles. 
	
	The standard spectral clustering method can be described as follows, where $ W \in \mathbb{R}^{n\times n} $ is the weight matrix associated with the spatial similarity graph and $k$ is the number of desired regions.
	
	\begin{itemize}
		\item Compute the normalized Laplacian matrix $L = D - W$, where $D$ is the diagonal matrix such that each element $D_{i,i} = \sum_{j=1 }^{n} W_{i,j}$.
		\item Compute the $k$ eigenvectors of the normalized Laplacian $D^{-1}L$ ($=I-D^{-1}W$) corresponding to the smallest $k$ eigenvalues.
		\item Apply the k-means clustering algorithm on the rows of the eigenvectors to obtain the final clusters.
	\end{itemize}
	
	The spectral clustering algorithm is very computationally expensive especially when we are dealing with a  large number of voxels $n$ 
	and a large number $k$ of regions. Therefore we develop high performance computing techniques to speed up the algorithm as described in the next section.

	\subsection{Parallel spectral clustering}
	A detailed description of our parallel spectral clustering algorithm and its application to a wide variety of datasets has been presented in 
	\cite{yu2016spectral}. Here we provide a high level overview of the implementation on a heterogeneous CPU-GPU platform. 
	\subsubsection{Parallel graph construction}
	We start by computing a sparse representation of the similarity graph on the CPU side using edge lists. Such representation is only computed once because all subjects are registered from the individual local space to the standard space, and hence will have the same patterns of spatial affinity. Each element $<i, j>$ on the edge list is a single edge that connects two nodes indexed by $i$ and $j$ within a spatial distance $r \leq 2$ . As the connectivity profiles are known, the computation of edge weights is a data-parallel task which is computed on the GPU with each thread executing on a small set of the edges.
	\begin{figure}[!hbt]
		\centering
		\includegraphics[width =\hsize]
		{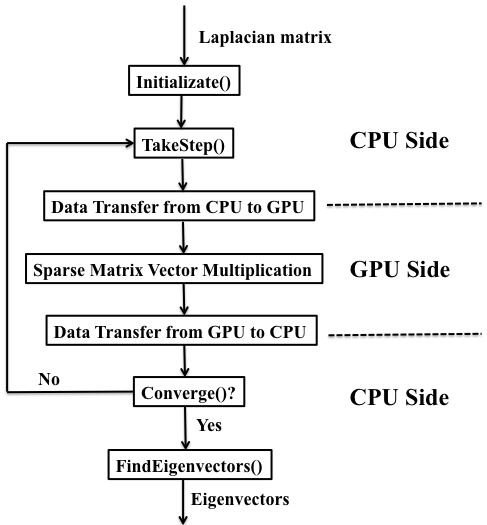}
		\caption{A fast implementation to solve the smallest $k$ eigenvalue problem}
		\label{eigensolver}
	\end{figure}
	
	\subsubsection{Parallel eigensolver}
	\par The most computational expensive part of the spectral clustering algorithm is the determination of the $k$ eigenvectors of a large real-valued  symmetric matrix corresponding to the smallest $k$ eigenvalues. 
	
	\par We combine an important feature of the ARPACK software named \emph{reverse communication interfaces} and the CUDA-based sparse matrix-vector multiplication. The reverse communication interfaces are a set of CPU-based interfaces that facilitate the process of solving large-scale eigenvalue problems. For each iteration, one of the interfaces \texttt{TakeStep()} provides a $n$-length vector used as the input, and the output of sparse matrix-vector multiplication is provided back to the interface \texttt{Converge()} to determine the convergence. In our implementation, the matrix-vector multiplication is performed on the GPU side using \emph{cuSPARSE} library. For each iteration, the input vector is transferred from the CPU to the GPU and the output vector is transfered back to the interface. The procedures of parallel eigensolver are described in Figure \ref{eigensolver}. A detailed description can be found in \cite{yu2016spectral}. 
	
	\subsubsection{Parallel k-means clustering}
	Modern scientific software packages and toolboxes such as Matlab\footnote{\href{http://www.mathworks.com/help/stats/kmeans.html}{http://www.mathworks.com/help/stats/kmeans.html}} and scikit-learn provide serial and parallel versions of the k-means clustering algorithm. However their CPU-based implementations are  inefficient for large-scale problems, especially when the number of clusters $k$ is very large. We briefly present a highly efficient implementation based on an open-source project\footnote{\href{https://github.com/bryancatanzaro/kmeans}{https://github.com/bryancatanzaro/kmeans}} which efficiently utilizes the Thrust and CUDA libraries and achieve very significant speedups. 
	
	\par The  eigenvectors corresponding to the smallest $k$ eigenvalues are represented as $ V \in \mathbb{R}^{n \times k}$, which initially reside on the CPU memory, where $n$ is the number of voxels and $k$ is the desired number of regions. The procedure is summarized in Algorithm 2.
	
	\begin{table}[t!]
		\centering
		\begin{tabular}{m{8cm}}
			\hline \hline
			\textbf{Algorithm 2} Parallel K-means Clustering Algorithm \\
			1. Transfer the eigenvectors $V \in \mathbb{R}^{n \times d}$ from the CPU to the GPU. \\
			2. Randomly select $k$ rows of $V$ as the centroids of the k clusters stored in $C \in \mathbb{R}^{k \times d}$\\
			3. While (the centroids change) do \\
			\hspace{0.2cm} Compute the pairwise distances $S \in \mathbb{R}^{n \times k} $ between rows of $V$ and the centroids.\\
			\hspace{0.2cm} Update the new label of each node. \\
			\hspace{0.2cm} Compute the new centroids of the clusters.\\
			4. Transfer the labeling result from GPU to CPU.\\
			\hline 
			\hline \hline	
		\end{tabular}
	\end{table}
	
	\subsection{Construction of Structural Brain Networks}\label{network}
	\par The brain  networks in our study are built based on the connectivity-based brain parcellations as well as the connectivity information revealed by probabilistic tractography. The network nodes correspond to the regions in the parcellation. We build and analyze structural brain networks at different scales which correspond to parcellations with different numbers $k$ of regions. 
	
	\par Given a parcellation whose regions define the nodes of the brain network, there is no consensus in the literature on how the edge 
	weights should be defined. At a high level, the edge weight  reflects the number of streamlines connecting the corresponding endpoints  \cite{zalesky2010whole, sun2015progressive, wang2016connectivity, iturria2013anatomical}, or the probability that the two endpoints can 
	be reached from each other \cite{gong2009age, kaiser2011tutorial}. In some cases, the weights are normalized considering the region volumes as well as the lengths of streamline trajectories. The key problem with defining edge weights is that we need  to accurately estimate the region-level connectivity strengths from the probabilistic tractography results that only involve voxel-level connectivity information. Note that the voxel-level connectivity information corresponds to an estimate of the number of streamlines connecting the two voxels, independently of the intermediate voxels. Many researchers such as \cite{gong2009age, herskovits2015edge} compute the region-level connections by accumulating the number of streamlines  between all voxels within the seed region and all voxels within the target region. Since some streamlines may pass through many voxels within the same region, such estimate will likely lead to a significant overestimation. Based on this observation, we introduce two notions of region-level connectivity strength $W(R_i, R_j)$ between two region $R_i$ and $R_j$. The first notion is 
	defined as follows.  
	
	\begin{equation}
	W(R_i, R_j) = \underset{v_{a} \in R_i, v_{b} \in R_j}{\max} W(v_a, v_b)
	\end{equation} 
	where $W(v_a, v_b)$ represents the number of streamlines connecting  the two voxels as generated by the tractographic results.
	
	\par It is easy to see that $W(R_i, R_j)$ is a lower bound of the total number of streamlines connecting  the corresponding regions. For parcellations with a large number of regions, we expect $W(R_i, R_j)$ to be reasonably close to the true value  since the number of voxels within each region will be relatively small. Overall we believe that this is a reasonable measure for the relative connectivity strength between regions.
	
	Our second definition normalizes the edge weights considering the strength of within-region connections. We first define the strength of the inter-region and within-region connections of $R_i$ and $R_j$ as follows.
	
	\begin{equation}
	C_{i, i} = \sum_{v_{a} \in R_i, v_{b} \in R_i} W(v_a, v_b)  
	\end{equation}
	
	\begin{equation}
	C_{j,j} = \sum_{v_{a} \in R_j, v_{b} \in R_j} W(v_a, v_b) 
	\end{equation}
	
	\begin{equation}
	C_{i, j} = \sum_{v_{a} \in R_i, v_{b} \in R_j} W(v_a, v_b) 
	\end{equation}
	
	\par The normalized edge weights are defined by
	
	\begin{equation}
	W(R_i, R_j) = \frac{C_{i, j}}{\sqrt{C_{i, i}C_{j, j}}}	
	\end{equation}
	By the Cauchy-Schwarz inequality, the normalized edge weights are always less or equal to 1. When most paths through region $R_i$ cross region $R_j$, the edge weight is close to 1. Otherwise when the intra-connections within regions are strong but the inter-connections between regions are weak, the probability is close to 0. 
	
	\par Since both definitions yield similar results in our subsequent analysis, we use the lower bound of the streamlines to define the edge weights of our structural brain networks for the rest of the paper.
	
	\subsection{Network Analysis}
	\subsubsection{Graph Theoretical Analysis}
	As described in section \ref{network}, the nodes of a structural brain network correspond to the brain regions defined by a parcellation, and the edge weights can be defined either by Equation (1) or (5). Since the results are similar with either definition of edge weights, we will adopt Equation (1) for the rest of the paper. In the next few sections, we present two types of network analysis. The first deals with the classical graph theoretic measures, and is coupled with an exploration of the relationship between the graph theoretical measures and the scale of brain networks. The study is conducted relative to two pairs of population groups that are expected to possess significant structural differences. The second type of analysis is based on spectral graph theory and a number of parameters derived from the spectra of structural brain networks.  We evaluate the robustness of the two types of network analysis in addition to exploring their statistical power relative to distinct population groups. 
	
	Many global and local graph-theoretic measures have been proposed to study different types of networks including social, biological, and social media networks.  Some of the studies that are related to our work appear in \cite{kaiser2011tutorial, rubinov2010complex}. Here we focus on a relatively few of the most frequently used graph theoretic measures  to study brain networks. 
	
	We note that the graph measures we are about to introduce are for unweighted and undirected graphs $U$ where $u_{i, j} = 1$ when there is an edge between nodes $i$ and $j$, and $u_{i, j} = 0$ otherwise. The number of nodes is $N$.
	
	\begin{itemize}
		\item Characteristic path length (CPL): The characteristic path length tries to capture the network integration and is computed as the average of the shortest paths between all pairs of vertices.
		\begin{equation}
		CPL = \frac{1}{N(N-1)}\sum_{i,j, i \neq j} d(i,j)
		\end{equation}
		where $d(i,j)$ is the shortest path between node $i$ and $j$.
		
		\item Global efficiency ($E_{global}$). The global efficiency of a graph is the average of the inverse of the shortest paths between all pairs of vertices and hence tries to capture how well  pairs of nodes are connected.  \cite{latora2001efficient}.
		\begin{equation}
		E_{global} = \frac{1}{N(N-1)}\sum_{i, j, i \neq j} \frac{1}{d(i,j)}
		\end{equation}
		
		\item Clustering coefficient. The clustering coefficient tries to capture graph separation and is defined as the average of the local clustering coefficient at each node.
		\begin{equation}
		C = \frac{1}{N}\sum C_i
		\end{equation}
		where the local clustering coefficient at node $i$ is defined as
		\begin{equation}
		C_i = \frac{2\Gamma_i}{deg_i(deg_i-1)}
		\end{equation}
		$\Gamma_i$ is the number of triangles around node $i$,
		\begin{equation}
		\Gamma_i = \frac{1}{2} \sum_{j,h} u_{i,j}u_{i,h}u_{j,h}
		\end{equation}
		Note that $deg_i >0$ in our case since isolated nodes are removed from the network.
		\item Sparsity ratio. The sparsity ratio is defined as the ratio of the number of edges over the number of possible edges between nodes, that is,
		\begin{equation}
		\text{Sparsity} = \frac{\sum_{i, j}u_{i, j}}{k(k-1)}
		\end{equation}	
	\end{itemize}
	
	\par A more detailed description of graph theoretic measures commonly used to analyze brain networks can be found in \cite{rubinov2010complex}. We preprocess our weighted structural brain networks into sparse unweighted graphs as follows \cite{kaiser2011tutorial},
	\begin{itemize}
		\item Remove self-loops whenever they exist.
		\item For each node, we want to eliminate edges with small weights. Since the total number of connections is significantly different across subjects, we first normalize the associated edge weights such that the sum of outgoing edge weights is equal to 1 and then eliminate the edges whose value are below a certain threshold $\epsilon$. In our study, we choose $\epsilon = 0.01$.
		\begin{equation}
		\label{normalize}
		w_{i, j} = \frac{W_{i,j}}{\sum_{j}{W_{i,j}}} 
		\end{equation}	
	\end{itemize}
	\subsubsection{Spectral Analysis} \label{spectral_analysis}
	We introduce some basic concepts from spectral graph theory, which we believe offers a more robust foundation to analyze and characterize brain networks. In essence, 
	spectral graph theory explores the fundamental relationship between graphs and the spectrum of matrices associated with these graphs, 
	typically either the adjacency matrix, or the Laplacian matrix. There are many tutorials and several textbooks that have been 
	written about spectral graph theory. See for example \cite{chung1997spectral}. 
	
	\par In our case, we focus on the normalized Laplacian matrix 
	whose eigenvalues are closely related to almost all major topological invariants of the graph. The normalized Laplacian of a 
	graph (or network) was introduced earlier in this paper when we gave an overview of the spectral clustering algorithm. It can be 
	defined using the following equation:
	
	\begin{equation}
	\label{laplacian}
	I - D^{-1}W
	\end{equation}
	where $W$ is the weight matrix of the graph and $D$ is the diagonal matrix such that each diagonal entry is defined by 
	$D_{i,i}=\sum_{j}W_{i,j}$. 
	
	\par The (normalized) spectrum of the graph is defined by the sequence of the eigenvalues of the normalized graph Laplacian ordered as follows:  $\lambda_1 =0 \leq \lambda_2 \leq ... \leq \lambda_n$. It can be shown that all eigenvalues are between 0 and 2 and that the largest eigenvalue $\lambda_n$ is equal to 2 if and only if the graph is bipartite \cite{chung1997spectral}. Moreover, the second smallest eigenvalue $\lambda_2$ is of special importance and has been  studied extensively in the literature. For example, it turns out that 
		$\lambda_2$ is related to the convergence rate of random walks on the graph, "chemical conductance," Cheeger's constant (roughly 
		the size of the sparset cut in the graph), graph diameter, and graph quasi-randomness \cite{chung1997spectral}. Other eigenvalues also play an important role 
		in capturing the graph structural properties. 
		
		\par Compared with traditional graph theoretic measures, spectral analysis seems to offer the following advantages,
		\begin{itemize}
			\item The eigenvalues of the graph Laplacian can be computed directly on the weighted structural brain network, which can be 
			shown to be robust to small noise. 
			\item The entire spectrum seems to capture almost all the invariant structural properties of the network.
		\end{itemize}
		
\par We will also illustrate the robustness of the graph spectrum relative to graph theoretic measures.
\begin{figure*}[!hbt]
	\centering
	\includegraphics[width =\hsize]{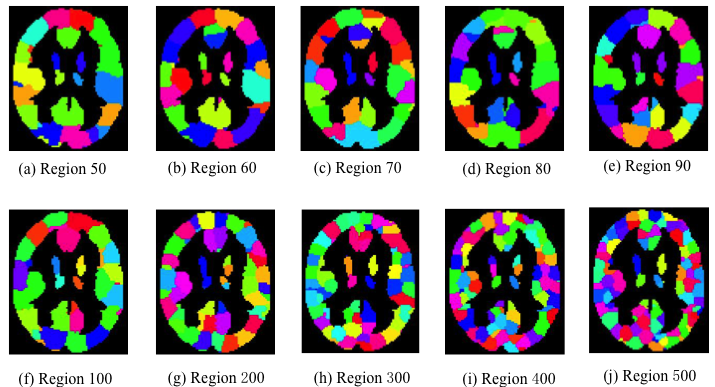}
	\caption{Example of connectivity-based brain parcellations at multiple scales. (Random coloring; axial view)}
	\label{parcellations}
\end{figure*}

\section{Results}
\subsection{Parcellation Evaluation}
\begin{figure*}[!hbt]
	\centering
	\includegraphics[width =\hsize]{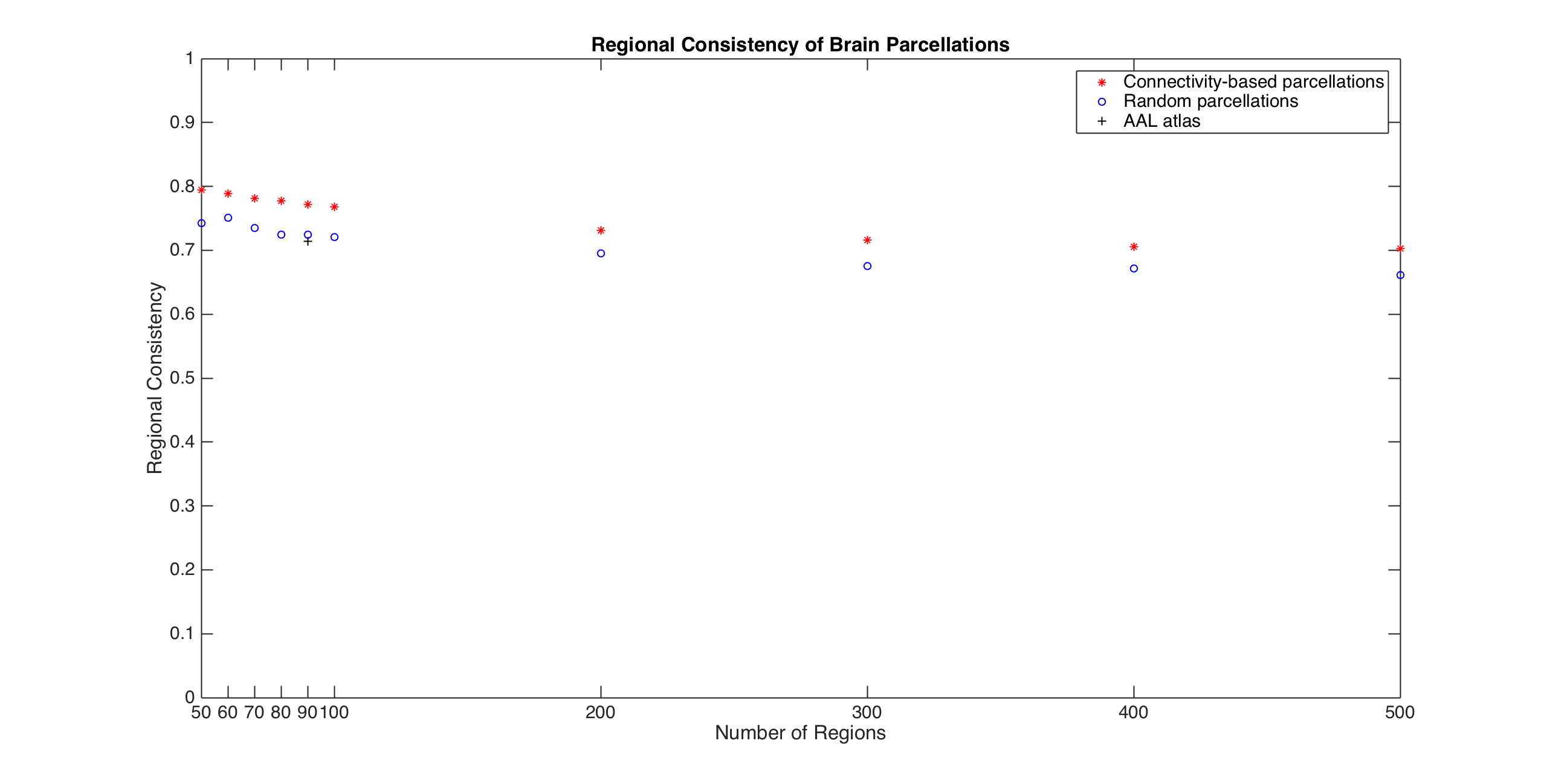}
	\caption{Regional consistency of connectivity-based brain parcellations and random parcellations.}
	\label{consistency_AAL}
\end{figure*}
	We start by shedding some light on the quality of our parcellations as the number of regions grows. The connectivity-based brain parcellations with different number of regions $k$ are shown in Figure \ref{parcellations}. We compare our connectivity-based parcellations against  randomly generated parcellations which are commonly used in many connectome studies. The randomly generated parcellations are a set of random spatial segmentations of AAL mask which are generated in a similar way as in \cite{zalesky2010whole}.

	\par We evaluate the \emph{regional consistency} of any parcellations as the average of the correlation coefficient of connectivity profiles between all pairs of voxels within the same region, which are defined in  equation~(\ref{consistency_equ}). Note that $P_{v_a}$ is the \emph{connectivity profile} of voxel $v_a$ defined by the parcellation and $|R_i|$ is the number of voxels within the region $R_i$. We justify this notion of connectivity profiles as follows. The goal of our pacellation is to build structural brain networks, and hence the voxels captured by a node should have very similar connectivity patterns to the rest of the nodes of the network. Otherwise we won't be able to compress all the voxels in a region into a single node. We also note that some voxels have no 
	or very few connections to other regions due to the nonlinear registration process from local space to standard 
	space; these voxels are eliminated before evaluating regional consistency. 
	
	\begin{equation}
	\label{consistency_equ}
	\frac{1}{k} \sum_{i = 1}^{k} \frac{1}{|R_i|(|R_i|-1)} \sum_{v_a, v_b\in C_i, v_b \neq v_a} \text{Corr}(P_{v_a}, P_{v_b}) 
	\end{equation} 
	
	\par The regional consistency measures the homogeneity in terms of the correlation coefficients 
	of the connectivity profiles averaged over all regions. Figures \ref{consistency_AAL} show the average regional consistency of the connectivity-based brain parcellations and random parcellations at multiple scales as well as AAL atlas over five subjects where the connectivity profiles are defined based on the corresponding parcellations. The initial segmentation used to define the connectivity profiles is the AAL atlas. For all subjects, the connectivity-based parcellations consistently possess higher structural regional consistency than the random parcellations ($p < 10^{-10}$) as well as AAL atlas, which suggests that connectivity-based brain parcellations have better voxel-wise homogeneity that are more suitable for constructing  structural brain networks. Moreover, the average regional consistency slightly decreases as the number of regions increases mainly because the connectivity profiles of fine-grained parcellations capture more connectivity information than the coarse-grained connectivity profiles. 
	
	\subsection{Network Analysis}
	\begin{table*}[hbt!]
		\caption{Group Information}
		\label{group}
		\centering
		\begin{tabular}{|c|c|c|c|c|}
			\hline
			& \multicolumn{2}{|c|}{\textbf{Comparison 1}} & \multicolumn{2}{|c|}{\textbf{Comparison 2}} \\
			\hline
			\textbf {Group Number}& \textbf{Group 1} & \textbf{Group 2} & \textbf{Group 1} & \textbf{Group 2} \\
			\hline
			\textbf{Age Range} & 4-10 & 60-80 & 11-20 & 60-80 \\
			\hline
			\textbf{Female} & 7 & 17 & 17 & 17\\
			\hline
			\textbf{Male} & 5 &  7 & 19 & 7\\
			\hline
			\textbf{Total} & 12 & 24 & 36 & 24\\
			\hline
		\end{tabular}
	\end{table*}
	\begin{table}[!hbt]
		\caption{Graph Measurements of Structural Brain Networks}
		\label{graph_measurements}
		\centering
		\begin{tabular}{|c|c|c|c|c|}
			\hline
			\textbf{k} & \textbf{CPL} & \textbf{$E_{global}$} & \textbf{C} & \textbf{Sparsity} \\
			\hline
			50 & 1.8641 & 0.6159 & 0.5253 & 0.2796\\
			\hline
			60 & 1.9610 & 0.5853 & 0.4917 & 0.2345\\
			\hline
			70 & 2.0375 & 0.5652 & 0.4711 & 0.2118\\
			\hline 
			80 & 2.0929 & 0.5488 & 0.4581 & 0.1892\\
			\hline
			90 & 2.1102 & 0.5394 & 0.4527 & 0.1708\\
			\hline
			100 & 2.2181 & 0.5182 & 0.4512 & 0.1560\\
			\hline
			200 & 2.6503 & 0.4326 & 0.4236 & 0.0819\\
			\hline
			300 & 2.9978 &0.3829 & 0.4268 & 0.0541\\
			\hline
			400 & 3.2075 &0.3562 &0.4214 & 0.0411\\
			\hline
			500 & 3.4294 & 0.3328 & 0.4209 & 0.0333\\
			\hline
		\end{tabular}
	\end{table}
	\subsubsection{Graph Theoretical Analysis}\label{graph_analysis}
		
	\par Table \ref{graph_measurements} shows the relationship between the graph theoretic measures introduced above and the number of regions. The global efficiency  decreases significantly with the number of regions while the characteristic path length increases significantly. In addition, the sparsity ratio decreases 
	quite significantly while the clustering coefficient decreases only very slightly with increasing $k$. These trends are consistent with prior results in the literature such as \cite{zalesky2010whole}. We will shed more light on the effect of 
	sparsity ratio on various graph-theoretic measures later in this paper. 
	
	\par We now examine the statistical differences of our graph theoretic measures between two sets of 
	pairs of population groups that are expected to possess different structural patterns. The two sets are defined in Table \ref{group}. Statistical differences are 
	measured using the p-values of the two-sample t-test, over a range of network sizes.

	\begin{table*}[!hbt]
		\caption{Discriminative Power of Graph Measurements (in terms of p-value of the two-sample $t$-test)}
		\label{comparison}
		\centering
		\begin{tabular}{|c|c|c|c|c|c|c|}
			\hline
			& \multicolumn{3}{|c|}{\textbf{Comparison 1}} & \multicolumn{3}{|c|}{\textbf{Comparison 2}} \\
			\hline
			\textbf{k} & \textbf{CPL} & \textbf{$E_{global}$} & \textbf{C} & \textbf{CPL} & \textbf{$E_{global}$} & \textbf{C}  \\
			\hline
			50 & $4.0466 \times 10^{-6}$ & $5.2804 \times 10^{-6}$ & 0.2965 & 0.0078 & 0.0038 & 0.8352\\
			\hline
			60 & $4.0429 \times 10^{-7}$ & $1.3503 \times 10^{-6}$ & 0.8713 & $7.8283 \times 10^{-4}$ & $5.6824 \times 10^{-4}$ & 0.6866\\
			\hline
			70 & $5.5281 \times 10^{-6}$ & $5.8380 \times 10^{-6}$ & 0.8108 & $5.6084 \times 10^{-4}$ & $6.7248 \times 10^{-4}$ & 0.2981\\
			\hline
			80 & $9.5953 \times 10^{-6}$ & $1.1765 \times 10^{-5}$ & 0.3231 & 0.0020 & 0.0031 & 0.4535\\
			\hline
			90 & $6.5179 \times 10^{-6}$ & $1.0460 \times 10^{-5}$ & 0.0934 & $5.1846 \times 10^{-4}$ & $7.2236 \times 10^{-4}$ & 0.5989\\
			\hline
			100 & $6.0881 \times 10^{-6}$ & $8.7734 \times 10^{-6}$ & 0.0351 & $3.3375 \times 10^{-4}$ & $5.1019 \times 10^{-4}$ & 0.2284 \\
			\hline
			200 & $3.0427 \times 10^{-7}$ & $4.8142 \times 10^{-7}$ & $7.7821 \times 10^{-7}$ & $2.2048 \times 10^{-4}$ & $8.7778 \times 10^{-4}$ & $3.4852 \times 10^{-4}$\\
			\hline
			300 & $5.3190 \times 10^{-5}$ & $7.0366 \times 10^{-5}$ & $3.2108 \times 10^{-8}$ & $6.5096 \times 10^{-4}$ & 0.0042 & $4.7756 \times 10^{-5}$ \\
			\hline
			400 & $4.7068 \times 10^{-5}$ & $3.0236 \times 10^{-5}$ & $1.2254 \times 10^{-6}$ & $7.0862 \times 10^{-4}$ & 0.0097 & $1.5504 \times 10^{-6}$ \\
			\hline
			500 & $1.4848 \times 10^{-5}$ & $2.1470 \times 10^{-5}$ & $9.5681 \times 10^{-6}$ & 0.0015 & 0.0204 & $6.1970 \times 10^{-5}$ \\
			\hline
		\end{tabular}
	\end{table*} 
	
	\begin{table*}[!hbt]
		\caption{Discriminative Power of Spectral Measurements (in terms of p-value from the two-sample $t$-test)}
		\label{spectral}
		\centering
		\begin{tabular}{|c|c|c|c|c|}
			\hline
			& \multicolumn{2}{|c|}{\textbf{Comparison 1}} & \multicolumn{2}{|c|}{\textbf{Comparison 2}} \\
			\hline
			
			\textbf{k} & \textbf{Modularity} & \textbf{$\lambda_2$} & \textbf{Modularity} & \textbf{$\lambda_2$} \\
			\hline
			50 & 0.6033 & 0.0012 &0.0146& 0.0094 \\
			\hline
			60 & 0.0223 & $1.9278 \times 10^{-4}$ & 0.0987 & $5.3686 \times 10^{-5}$ \\
			\hline
			70 & 0.0049 & $1.1783 \times 10^{-4}$ &  0.0282 & $2.3541 \times 10^{-4}$ \\
			\hline
			80 & 0.0014 & $1.0335 \times 10^{-4}$ &  0.0117 & $8.0141 \times 10^{-5}$\\
			\hline
			90 & 0.0010 & $1.6525 \times 10^{-4}$ &  0.0238 & $3.7983 \times 10^{-4}$ \\
			\hline
			100 & 0.0024 & $1.1762 \times 10^{-4}$ &  0.0026 & $5.7496 \times 10^{-5}$ \\
			\hline
			200 & $7.6224 \times 10^{-5}$ & $1.0702 \times 10^{-6}$ & 0.0281 & $9.0932 \times 10^{-6}$ \\
			\hline
			300 & $1.0352 \times 10^{-4}$ & $1.5826 \times 10^{-6}$  & 0.0275 & $3.3413 \times 10^{-4}$ \\
			\hline
			400 & $4.6564 \times 10^{-4}$ & $3.1102 \times 10^{-6}$  & 0.0439 & $3.2788 \times 10^{-4}$ \\
			\hline
			500 & $3.1176 \times 10^{-4}$ & $4.5524 \times 10^{-6}$ & 0.1043 & 0.0013 \\
			\hline
		\end{tabular}
	\end{table*} 
	
	\begin{table*}[!hbt]
		\caption{Prediction Accuracy with Graph Topological Measurements and Spectral Properties of Structural Brain Networks}
		\label{Accuracy}
		\centering
		\begin{tabular}{|c|c|c|c|c|}
			\hline
			& \multicolumn{2}{|c|}{\textbf{k = 500}} & \multicolumn{2}{|c|}{\textbf{k = 90}} \\
			\hline
			\textbf{Feature selection} & \textbf{Accuracy (Group 1)} & \textbf{Accuracy (Group 2)} & \textbf{Accuracy (Group 1)} & \textbf{Accuracy (Group 2)} \\
			\hline
			CPL+$E_{global}$ + C & 0.8056 & 0.7167 & 0.7778 & 0.6500\\
			\hline
			$\lambda_2$ & 0.8056 & 0.7333 & 0.7778 & 0.7000\\
			\hline
			$\lambda_2$ + $\lambda_3$ & 0.8333 & 0.7667 & 0.7778 & 0.7333\\
			\hline
		\end{tabular}
	\end{table*}
	
	\par Table \ref{comparison} includes the p-values for each group pair and when the number of nodes varies from $k=50$ up to $k=500$.

	\par We note that the characteristic path length and global efficiency are statistically different 
	for each pair of population groups, regardless of the number of nodes in the network. However 
	the clustering coefficient becomes statistically discriminative for each pair of the population groups 
	only when the number of regions becomes relatively large, at least $k\geq 200$. This illustrates the 
	fact that larger brain networks can capture more detailed structural differences than smaller networks.
	\par Since the graph-theoretic measures represent discrete aggregated network measurements, they lack 
	the ability to capture network characteristics in detail, which motivates the analysis based on the entire 
	spectrum of the network of each subject as described next. 

	\subsubsection{Spectral Analysis}	
	\begin{figure*}[!hbt]
		\centering
		\begin{minipage}{.4\textwidth}
			\centering
			\includegraphics[width=1.\hsize]{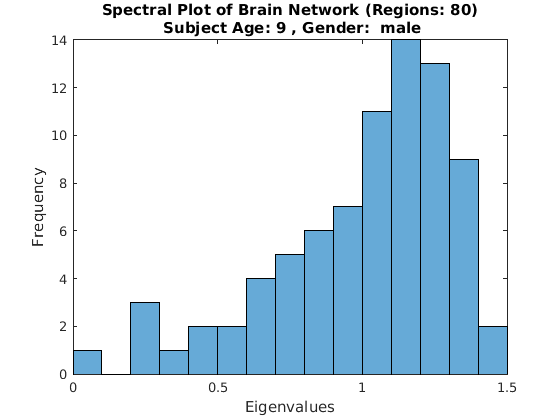}
			\caption{Spectral Plot of Structural Brain Network.(Age: 9, Region 80, $\lambda_2 = 0.2013$)}
			\label{9_80}
		\end{minipage}%
		\begin{minipage}{.4\textwidth}
			\centering
			\includegraphics[width=1.\hsize]{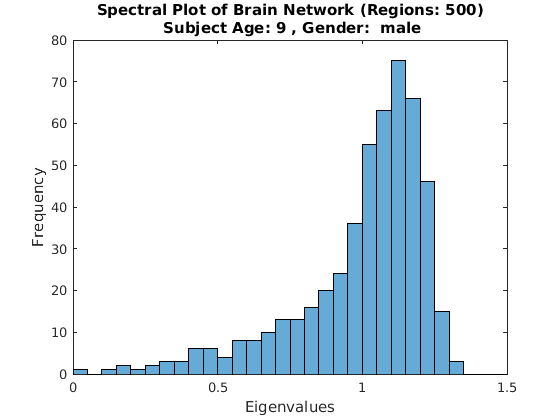}
			\caption{Spectral Plot of Structural Brain Network.(Age: 9, Region 500, $\lambda_2 = 0.1323$)}
			\label{9_500}
		\end{minipage}
		
		\centering
		\begin{minipage}{.4\textwidth}
			\centering
			\includegraphics[width=1.\hsize]{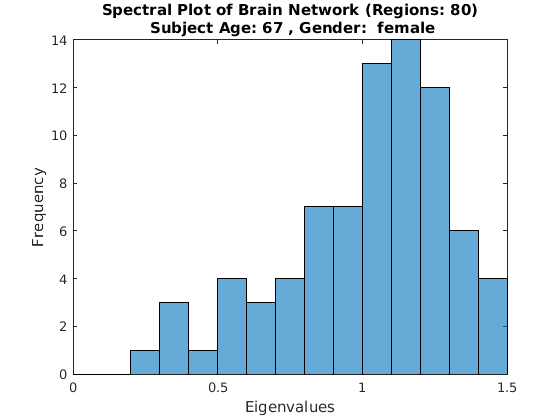}
			\caption{Spectral Plot of Structural Brain Network. (Age: 67, Region 80, $\lambda_2 = 0.2662$)}
			\label{67_80}
		\end{minipage}%
		\begin{minipage}{.4\textwidth}
			\centering
			\includegraphics[width=1.\hsize]{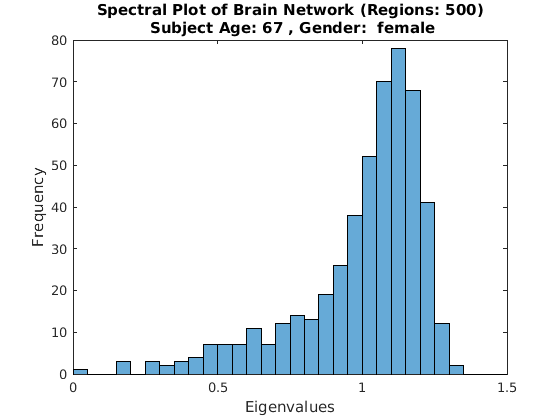}
			\caption{Spectral Plot of Structural Brain Network. (Age: 67, Region 500, $\lambda_2 = 0.1506$)}
			\label{67_500}
		\end{minipage}
	\end{figure*}

	\par Figures \ref{9_80} -\ref{67_500} show the spectral plots (histograms of the normalized Laplacian eigenvalues) of the weighted structural brain networks of two subjects, each with two distinct values of the number of nodes. Note that the spectral plots share very similar shapes, which hold over the subjects in our data set especially for larger networks. Note also that the second 
	eigenvalue decreases with the size of the network and larger fractions of smaller eigenvalues appear with larger networks (which 
	indicate more modular structures). Note also that the spectrum of larger networks is further and further 
	away from $2$, which implies that the larger networks are further and further away from being bipartite.

	We note that spectral analysis has been recently used to analyze different types of networks such as social networks and biological networks \cite{banerjee2008spectrum}. Most related to our work are the unweighted  neural networks of macaque, cat and Caenorhabditis elegans \cite{de2014laplacian}. 
	
	\par We introduce two global parameters based on the spectrum of a network.  
	\begin{itemize}
		\item The second smallest eigenvalue $\lambda_2$. As mentioned before, this parameter seems to capture a number of critical 
		topological properties of the network.
		\item Modularity. The number of small eigenvalues in the spectral plot reflects the extent of structural modularity in 
		the network. In this paper, we define the modularity as the number of eigenvalues less than a certain threshold, say $\gamma = 0.3$. 
		
	\end{itemize}
	
	\par Using the parameters introduced above, we explore their statistical variations for the two comparison groups as a function of the number of nodes in the network. Table \ref{spectral} shows the p-values of the samples of the spectral parameters corresponding to the two sets of population groups defined  in Table \ref{group}. It is easy to see that the  modularity parameter and second eigenvalue are  significantly different for each set of population groups, especially for large-scale structural brain networks. 
	These findings suggest that fine-grained brain networks are better 
	able at capturing subtle structural patterns than coarser-grain networks.

	\subsubsection{Robustness of Spectral Graph Parameters}
	In this section we shed some light on the relative robustness of the two types of network analysis. It is well-known that the graph 
	theoretic measures depend heavily on the preprocessing step.  Recall that the preprocessing step  eliminates edges with very weak connections and transforms relatively dense graphs into sparse graphs. Graph measures such as the characteristic path length and clustering coefficient  are sensitive to  the \emph{sparsity ratio} defined by the fraction of non-zero entries in the adjacent matrix, which is controlled by the threshold value used during the preprocessing step. However, the Laplacian spectrum seems to be more robust to the threshold value chosen to control the sparsity of brain networks. Table \ref{robust1} lists the values of some graph-theoretic measures of a 500-node brain network using different threshold values. As can be seen from these results, the values of the graph theoretic parameters change substantially as a function of 
	the threshold value, which cast some doubt on their utility to analyze brain networks. 
	
	\begin{table}[hbt!]
		\caption{Robustness of Graph Measurements with Different Threshold Value}
		\label{robust1}
		\centering
		\begin{tabular}{|c|c|c|c|c|}
			\hline
			\textbf{$\epsilon$} & \textbf{CPL} &  \textbf{$E_{global}$} & \textbf{C} & \textbf{Sparsity ratio} \\
			\hline
			0 &  1.2819 & 0.8590 & 0.8185 & 0.7181\\
			\hline
			0.0001 & 1.4128 & 0.7936 & 0.7075 & 0.5873 \\
			\hline
			0.001 & 2.0083 & 0.5499 & 0.4294 & 0.1538\\
			\hline
			0.01 &  3.4294 & 0.3328 & 0.4209 & 0.0333\\
			\hline
		\end{tabular}
	\end{table}
	\begin{figure}[!t]
		\centering
		\label{spectral_plot_threshold}
		\includegraphics[width =\hsize]{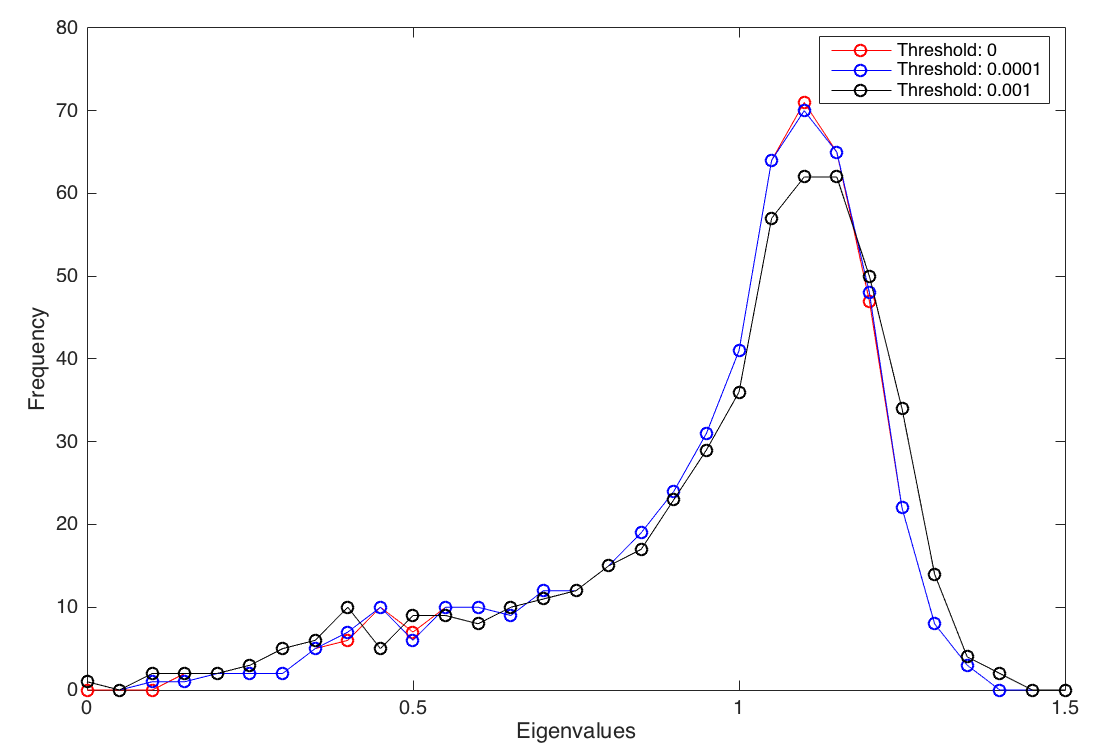}
		\caption{Spectral Plot of Brain Network (Regions: 500), Subject Age: 41, Gender: male}
	\end{figure}

	\par  Figure 13 shows the spectral plots corresponding to the weighted brain networks using the same set of threshold values as before. The spectral plots are similar despite that the fact that the sparsity ratios are significantly different. These results indicate that the spectral properties of the high-resolution brain networks are more robust  than traditional graph theoretic measures relative to the threshold values using 
	during preprocessing.

	\subsubsection{Classification based on Network Parameters}
	Tables \ref{comparison} and \ref{spectral} illustrated the increasingly statistical differences of graph measures for our distinct population groups as we increase the number of nodes. We provide here an alternative evidence by building classifiers for each of our pair of population groups and measure their accuracy relative to the number of nodes in the network.  We train a Support Vector Machine (SVM) classifier with linear kernel\footnote{\href{http://www.mathworks.com/help/stats/svmtrain.html}{http://www.mathworks.com/help/stats/svmtrain.html}} on the two sets of subjects of Comparison~1 and Comparison~2 defined in table~\ref{group} using a variety of graph features. To measure the 
	classification accuracy, we use the Leave-one-out cross-validation (LOOCV) on both sets of population groups. Table~\ref{Accuracy} shows the classification accuracy based onthe use of different graph theoretic and graph spectral features. 
	
	\par For the fine-grained brain networks (k = 500), the SVM classifier achieve consistently better accuracy than using coarse-grain structural 
	brain networks. Note that the relative high accuracy achieved by using only the second eigenvalue of the normalized Laplacian of the network. 
	
	\subsection{Computational Efficiency}
	\begin{table}[t!]
		\caption{Problem size}
		\label{problem}
		\centering
		\begin{tabular}{|c|c|c|c|}
			\hline
			\# of voxels & \# of edges & \# of clusters\\
			\hline
			142541 & 3992290 & 500 \\ 
			\hline 
		\end{tabular}
	\end{table}
	\par In this section, we briefly provide an indication of the  time efficiency of our parallel implementation of the connectivity based brain parcellation scheme. We compare our fast implementation to the best multithreaded implementations we could achieve using Matlab and Python scientific packages. After removing  isolated voxels of the data of an arbitrary subject, we obtain a connectivity graph characterized  in Table \ref{problem}. 
	
	\par
\begin{figure}[!t]
		\centering
		\label{DTI}
		\includegraphics[width = .8\hsize]{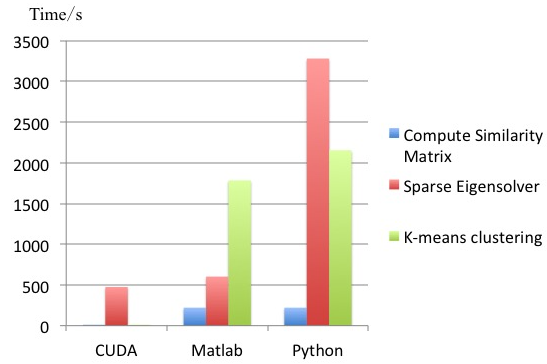}
		\caption{Time costs of brain parcellation scheme for one iteration}
	\end{figure}

 Figure \ref{DTI} shows the time costs (in seconds) of each major step for a single iteration of our parcellation algorithm. Note that our 
	algorithm typically converges in less than 4 iterations. Notice that the construction of the similarity matrix and the execution of the  k-means algorithm are substantially accelerated, which leads to about 5x speedup of the overall execution time compared to the fast Matlab implementation. We also obtain a good speedup for computing the $k$ eigenvalues and eigenvectors making use of very fast  sparse matrix-vector multiplication. On our CPU-GPU system, the  overall execution time of each iteration is reduced to less than 10 minutes for 
	$k=500$, which makes it quite 
	feasible to generate almost arbitrarily large size parcellations.
	
	\section{Conclusion}
	In this paper, we presented a parallel connectivity-based whole-brain parcellation scheme that was implemented on a heterogeneous CPU-GPU platform. The resulting algorithm generates large-scale parcellations quickly with provably structurally homogeneous regions. The algorithm was used to build structural brain networks with varying numbers of nodes up to $500$ nodes. These brain networks were then analyzed using 
	traditional graph theoretic measures as well as new measures that are based on spectral graph theory. Using data from populatioin groups that are expected to have significantly different structural patterns, we illustrated the increasing statistical power of the neworks as the number 
	of nodes grows. Moreover, we provided preliminary evidence on the robustness of using spectral parameters to analyze brain networks and 
	their potential for discriminating between different population groups. We are currently persuing this line of research trying to shed more 
	light on the use of spectral graph theory to characterize brain networks for different population groups.

	
	\section{Acknowledgment}
	We gratefully acknowledge funding provided by The University of Maryland/Mpowering the State through the Center for Health-related Informatics and Bioimaging (CHIB) and the NVIDIA Research Excellence
	Center at the University of Maryland.

	\bibliography{mybibfile}
\end{document}